# Low-temperature hysteresis broadening emerging from domain-wall creep dynamics in a two-phase competing system

Keisuke Matsuura[1,*], Yo Nishizawa[2], Yuto Kinoshita[3], Takashi Kurumaji[4], Atsushi Miyake[3], Hiroshi Oike[1,2,5], Masashi Tokunaga[1,3], Yoshinori Tokura[1,2,6] and Fumitaka Kagawa[1,2,7,†]

[1] *RIKEN Center for Emergent Matter Science, Wako 351-0198, Japan*
[2] *Department of Applied Physics and Quantum-Phase Electronics Center (QPEC), University of Tokyo, Tokyo 113-8656, Japan*
[3] *The Institute for Solid State Physics, The University of Tokyo, Kashiwa, Chiba 277-8581, Japan*
[4] *Department of Advanced Materials Science, The University of Tokyo, Kashiwa 277-8561, Japan*
[5] *PRESTO, Japan Science and Technology Agency (JST), Kawaguchi 332-0012, Japan*
[6] *Tokyo College, University of Tokyo, Tokyo 113-8656, Japan*
[7] *Department of Physics, Tokyo Institute of Technology, Tokyo 152-8551, Japan*

*Present address: Department of Physics, Tokyo Institute of Technology, Tokyo 152-8551, Japan
email: *matsuura@phys.titech.ac.jp (K.M.); † kagawa@phys.titech.ac.jp (F.K.)

**Abstract**

Hysteretic behaviour accompanies any first-order phase transition, forming a basis for many applications. However, its quantitative understanding remains challenging, and even a qualitative understanding of pronounced hysteresis broadening at low temperature, which is often observed in magnetic-field-induced first-order phase transition materials, is unclear. Here, we show that such pronounced hysteresis broadening emerges if the phase-front velocity during the first-order phase transition exhibits an activated behaviour as a function of both temperature and magnetic field. This is demonstrated by using real-space magnetic imaging techniques, for the magnetic-field-induced first-order phase transition between antiferromagnetic and ferrimagnetic phases in $(Fe_{0.95}Zn_{0.05})_2Mo_3O_8$. When combined with the Kolmogorov-Avrami-Ishibashi model, the observed activated temperature- and field-dependences of the growth velocity of the emerging antiferromagnetic domain quantitatively reproduce the pronounced hysteresis broadening. Furthermore, the same approach also reproduces the field-sweep-rate dependence of the transition field observed in the experiment. Our findings thus provide a quantitative and comprehensive understanding of pronounced hysteresis broadening from the microscopic perspective of domain growth.

**Introduction**

Functionalities carried by ferroic materials are closely related to their hysteretic behaviour that accompanies the first-order phase transition (FOT) associated with the order parameter reversal. Ferromagnets or ferroelectric materials with a high coercive magnetic or electric field facilitate a remnant order parameter; these materials have been applied to permanent magnets[1], piezo elements[2] and recording media[3,4]. Alternatively, those materials with a low coercive field have superior sensitivity and switching ability[5] have been applied to motors, capacitors, transformers, and power supplies. In addition to these commercial products, the hysteretic behaviour accompanying a field-induced FOT between competing phases with different symmetries plays a key role in exotic functionalities, such as nonvolatile resistance control by a magnetic field[6,7] and magneto- or electrocaloric refrigeration[8]. From a microscopic viewpoint, hysteretic behaviour accompanying an FOT originates from the nonequilibrium evolution dynamics of the emerging phase, such as nucleation and growth; thus, the hysteretic behaviour tends to vary with the sweep rate of relevant parameters, also dictating the performance under high-speed operation[9].

For a field-induced FOT in a single-crystalline sample, the transition profile is usually sharp, and accordingly, two characteristic transition fields, $H^*_{high}$ and $H^*_{low}$, can be defined: one is higher than the field of the equilibrium FOT, $H_c$, and the other is lower. Thus, by plotting $H^*_{high}$ and $H^*_{low}$ at various temperatures, two hysteresis lines, $H^*_{high}(T)$ and $H^*_{low}(T)$, sandwiching the equilibrium FOT line, $H_c(T)$, can be drawn in the phase diagram. However, in some materials exhibiting a magnetic-field-induced FOT, the difference between $H^*_{high}(T)$ and $H^*_{low}(T)$, i.e., the hysteresis width, starts to increase below a certain temperature and eventually shows a large hysteresis width of more than several Tesla at low temperatures. For example, the hysteresis width between ferrimagnetic up and down states in multiferroic $LuFe_2O_4$ reaches ~18 T at 4.2 K [10]. Although it is rare for ferroic materials to exhibit such high coercivity, there are many examples of field-induced FOTs between competing phases of different symmetries that exhibit pronounced hysteresis broadening toward low temperatures (Fig. 1**a**), for instance, the colossal magnetoresistive manganite[6], $Gd_5Ge_4$[11], $LaFe_{12}B_6$[12], doped $CeFe_2$ alloy[13], and doped $Mn_2Sb$[14,15]. Similar hysteresis broadening has also been observed for a structural FOT in martensitic materials[16,17] and aqueous solutions[18], implying that common physics, which is not limited to magnetic materials, underlies these observations. However, different equations have been used for different systems to analyze hysteresis lines[10,16,17,19-21], and thus, the universal relationship between microscopic domain-wall motion and the broadening of hysteresis lines at low temperature remains unclear. Such hysteresis broadening is also potentially relevant to thermal-quenching-induced metastable states in quantum materials[22].

For example, IrTe$_2$ exhibits a metastable superconducting state when thermal quenching is applied[23]. This behaviour can be understood by assuming the hysteresis broadening accompanying the chemical-doping-induced FOT, although it is not possible to demonstrate directly the hypothetical broadening in this system by continuously changing the composition at low temperatures.

Among materials exhibiting low-temperature hysteresis broadening, we targeted (Fe$_{0.95}$Zn$_{0.05}$)$_2$Mo$_3$O$_8$ (Fig.1**b**) because the hysteretic magnetic-field-induced FOT is accessible with a moderate magnetic field[24,25], as shown in the phase diagram (Figs. 1**c,d**); the FOT nature is confirmed by the fact that the extensive variables such as entropy and magnetization are discontinuous between the antiferromagnetic (AFM) and ferrimagnetic (FRI) phases and no critical phenomena are observed[32]. In (Fe$_{1-y}$Zn$_y$)$_2$Mo$_3$O$_8$, the magnetism originates from two crystallographically inequivalent Fe$^{2+}$ (I) and Fe$^{2+}$(II) sites (Fig.1**b**), and Mo$^{4+}$ layers have no contribution to the magnetism[26]. The magnetic moments of Fe(II) sites are slightly greater than those of Fe(I) sites[24]. In Fe$_2$Mo$_3$O$_8$, a metamagnetic transition occurs under magnetic fields along the *c*-axis[25,27]. In the low-field region, the two Fe(I) sites are antiferromagnetically ordered, and the same is true for the two Fe(II) sites, thus resulting in an antiferromagnetic (AFM) phase with no net macroscopic magnetization. When a sufficiently high magnetic field (~7 T) is applied, by contrast, the ferromagnetically aligned Fe(I) sites and the ferromagnetically aligned Fe(II) sites are antiferromagnetically ordered, thus resulting in a ferrimagnetic (FRI) phase with macroscopic magnetization. Doped Zn ions selectively occupy the Fe(I)-sites, which decreases the critical field of the metamagnetic transition[28] and causes the broadening of the hysteresis region at low temperatures[25]. For compositions above y=0.05, the FRI phase is so stable that the FRI to AFM transition can be observed only in a very narrow temperature-field region; on the other hand, for compositions below y=0.05, the AFM phase is so stable that the magnetic-field-induced FOT is difficult to observe. For these reasons, the present study focused on the composition of y=0.05, in which the transition field is well accessible and thus experiments are feasible.

In this Article, we show that the pronounced hysteresis broadening originates from an activated behaviour of domain-wall dynamics as a function of both temperature and magnetic field, i.e., the creep motion of a domain wall. This is evidenced by real-space magnetic imaging experiments and numerical simulations. Furthermore, we also find that a sweep-rate dependent magnetic hysteresis loop, which is important for understanding the performance under high-speed operation, is also well reproduced within our numerical simulations. The present study unveils a strong correlation between microscopic domain-wall motion and the macroscopic transition accompanied by low-temperature hysteresis broadening.

## Results and discussion

### Observation of an antiferromagnetic-domain growth.

To reveal the microscopic origin of the hysteresis broadening in $(Fe_{0.95}Zn_{0.05})_2Mo_3O_8$ from the perspective of the nonequilibrium phase-evolution dynamics, we performed magnetic force microscopy (MFM) and magneto-optical Kerr effect (MOKE) imaging. In particular, we focused on AFM phase evolution in the matrix of the FRI phase, close to the lower-field boundary of the hysteresis region. To this end, a single-phase FRI state was first prepared by applying a magnetic field of 7 T at high temperature (~25 K), followed by field-cooling to a target temperature under 7 T. Then, the magnetic field was decreased to a target field, and the time evolution of the emergent AFM domain was measured while holding the temperature and magnetic field constant. Figures 1**e-g** show the time evolution of the MFM images at 15 K and 1.8 T, noting that this comes from a magnetic signal (Supplementary Note 1). In the initial state, a small disk-shaped AFM domain was observed in the matrix of the FRI phase, and it gradually and continuously expanded over ~$10^4$ s (Figs. 1**f,g**). A similar AFM domain was also observed at a different position. While MFM has a high spatial resolution of < 50 nm, image acquisition takes 10–20 minutes, and thus, rapid domain growth cannot be tracked. To track the faster motion of the AFM-domain growth, we utilized MOKE imaging, which has a lower spatial resolution, ~1 μm, but a faster image acquisition time, i.e., 1 s. In MOKE imaging, the AFM domains were observed as similar multiple circular spots with dark boundaries in a much wider field of view (Fig. 1**h**). Note that images shown in panels Figs. 1**h** and 1**i-k** are divided by the raw MOKE image at 0 s to remove the influence of the position-dependent intensity of the incident light. Figures 1**i-k** display the AFM domain growth at 15 K and 0.9 T, demonstrating that compared with the data at a higher magnetic field (Figs. 1**e-g**), the growth speed of the AFM domains is pronouncedly faster.

We found that the radius of the circular AFM domain changes almost linearly with respect to elapsed time, as shown in Figs. 1**l** and 1**m**, and from this behaviour, a constant growth velocity, $v$, of the AFM domain was defined as the slope of this plot; the time evolution of the circular domain appears monotonic, suggesting that the spatial distribution of impurities can be regarded as homogenous as far as the present domain-wall dynamics are concerned. The same analysis was performed for various temperatures, $T$, and magnetic fields, $H$, and thus we were successful in deriving $v(H, T)$ for $T \leq 20$ K: Above this temperature, the image contrast significantly decreased, although the reason is not clear; therefore, the analysis was not possible. The results are summarized as a contour plot of the growth velocity in Fig. 1**n** (see also Supplementary Note 2 for the reproducibility at different positions), and to gain insight into the relationship between $v(H, T)$ and the hysteresis line $H^*_{\text{low}}(T)$, the contour plot is superimposed on the phase diagram. Here, two important points can be highlighted. On the one hand, the curvature of the contour line is similar to that of the $H^*_{\text{low}}(T)$ line, suggesting that the profile of the hysteresis line is closely related to the microscopic growth process of the AFM domain. On the other hand, although the growth speed near the $H^*_{\text{low}}(T)$ line is expected to have a particularly important impact on the macroscopic phase evolution under the continuous field-sweeping experiments, the information was not obtained from the above analysis because the growth speed is far beyond the detection limit, ~1 μm s$^{-1}$.

**Creep motion of a domain wall between the antiferromagnetic and ferrimagnetic phases.**

To obtain $v(H, T)$ near the $H^*_{low}(T)$ line, the microscopic mechanism should be identified together with the corresponding formula. To this end, below we consider the application of the so-called modified Merz's law that describes the creep motion of an elastic interface, such as a domain wall (DW), in a disordered medium[29,30,31]. The characteristic of the modified Merz's law is that the velocity of a moving DW obeys activated behaviour with respect to both temperature, $T$, and the driving force exerted on the DW, $F$. The equation is given as $v(T, F) = A*\exp[-\Delta/(TF^\mu)]$, where $\mu$ is a so-called dynamical exponent and $A$ and $\Delta$ represent the limit of high speed and temperature-force-composite activation barrier, respectively. When applying this equation to $(Fe_{1-y}Zn_y)_2Mo_3O_8$, the driving force should be measured with reference to $H_c(T)$ (for more details, see Supplementary Note 3), which has recently been determined by making full use of the thermodynamic relations[32] as indicated by white circles in Fig. 1**d**. Thus, in the present case, the modified Merz's law is given as

$$v(T, \delta H) = A \exp\left[-\frac{\Delta}{T \times (\delta H)^\mu}\right], \tag{1}$$

where $\delta H(T) = |H(T) - H_c(T)|$. Equation (1) describes the average velocity of the creep motion, rather than the instantaneous velocity that may be affected by the details of the spatial distribution of impurities.

Figure 2**a** displays the growth velocities at various temperatures, plotted against $1/(\delta H)^\mu$. We find that the observed velocities are well reproduced by equation (1) (solid lines, Fig. 2**a)**, indicating that the observed DW motion should be dictated by the creep motion. We note, however, that it is not so obvious to what extent the creep motion describes the actual DW dynamics outside the examined parameter region; in fact, previous theoretical studies predict that the DW dynamics change from the creep to flow regimes as the driving force increases [30]. We will discuss this issue later. The optimized adjustable parameters are $A$=3806.5 [m s$^{-1}$], which is quite close to the expected velocity of the acoustic phonon in $Fe_2Mo_3O_8$, 2.7–6.6×10$^3$ m s$^{-1}$ calculated from the elastic constants[33], and $\mu$=0.4. The similarity between the value of $A$ and the phonon velocity may imply that structural change accompanies the metamagnetic transition and thus affects the dynamics of DW. The value of Δ is nearly constant (~500) within the measured temperature range (Fig. 2**b**), and the residual weak temperature dependence may be attributed to that of the order parameter. In ferroic materials, it is known that the value of $\mu$ depends on both the dimensionality of the DW and the type of disorder, which is generally classified into random-bond or random-field types[30,34,35]. However, for the case of DWs separating distinct symmetry phases, such as AFM and FRI, such classification of disorder is not straightforward (for details, see Supplementary Note 4). Thus, although $\mu$ = 0.4 appears to be close to $\mu$ = 0.5–0.6, which is a value reported for two-dimensional DWs under the influence of random-bond-type disorder[31,36], its implications are not clear.

**Origin of the pronounced hysteresis broadening.**

Nevertheless, having established that equation (1) reproduces the observed $v(T, \delta H)$ well, we can supplement the data near the $H^*_{\text{low}}(T)$ line by the extrapolation according to equation (1), enabling us to quantitatively investigate the relationship between the creep-like growth velocity and the pronounced hysteresis broadening. To this end, we use the Kolmogorov-Avrami-Ishibashi (KAI) model[37,38], which can calculate the volume fraction of a growing domain for a given DW velocity while avoiding the complexity caused by domain coalescence (Fig. 3**a**). Given that in the present observation, the nucleation sites are always at the same positions, we applied the formalism of the KAI model with the so-called one-step nucleation mechanism (for details, see Supplementary Note 5):

$$c(t) = 1 - \exp[-NS(t)], \qquad (2)$$

where $c(t)$, $N$, and $S(t)$ are the time-dependent volume fraction of the AFM domain, the density of nucleation sites, and the time ($t$)-dependent extended area caused by a single nucleation site, respectively. When calculating $S(t)$, the DW velocity should be integrated with respect to $t$: In the present study, because a phase transition under continuous field-sweeping experiments is considered, the DW velocity is time dependent via the time-dependent driving force, $\delta H(t)$. Thus, $c(t)$ is formulated as follows:

$$c(t) = 1 - \exp\left[-N \times \alpha(n) \times \left\{\int_0^t v(T, \delta H(s))\, ds\right\}^n\right], \tag{3}$$

where $\alpha(n)$ is a shape factor and equal to 2, $\pi$, or $4\pi/3$ depending on stripe-domain growth (one dimensional, $n=1$), circular-domain growth (two dimensional, $n=2$), or spherical-domain growth (three dimensional, $n=3$), respectively.

The calculation of the phase evolution was performed by using the experimentally obtained parameter set, $A=3806.5$, $\mu=0.4$ and $\Delta=500$. We define the transition field $\delta H_c(T)$ as the field at which the phase evolution is half complete. The experimental transition profile at 20 K and the corresponding calculated results are shown in Figs. 3**b** and 3**c**, respectively. The details of the transition profile, such as the value of $\delta H_c(T)$ and the sharpness of the transition, depend on $N$ and $n$ (Fig. 3**c**). However, we find that the shape of the hysteresis line is not sensitive to $N$ and $n$ (Fig. 3**d** and Supplementary Figs. 4**a**,**b**). Therefore, upon comparing the hysteresis line profiles of the calculation and experiments, the normalized temperature dependence of $\delta H_c(T)/\delta H_c(12\,\text{K})$ is the quantity that should be considered. The comparison is displayed in Fig. 3**d**, showing good agreement with the experiment. Furthermore, the profile of the hysteresis broadening is less pronounced for a higher $\mu$ value, such as $\mu = 1$ (Supplementary Note 5), which is a value reported for DWs in ferroic materials with random-field type disorder[31,39,40]. These observations indicate that the peculiar $T$ and $\delta H$ dependences of the DW creep velocity, underpinned by a low $\mu$ value, are at the heart of the pronounced hysteresis broadening.

**Sweep-rate dependence.**
Finally, to further corroborate the relationship between the creep dynamics of the DWs and the pronounced hysteresis broadening, it would be important to test another aspect of the hysteresis line, that is, the sweep-rate dependence. If the $H^*_{\text{low}}(T)$ line is dictated by the creep dynamics, the $H^*_{\text{low}}(T)$ line should depend on the sweep rate as a natural consequence of the kinetic aspect. Figure 4**a** shows the sweep rate dependences of the isothermal magnetization curves at 20 K, and it can be seen that the FRI-AFM transition field (white circles) appreciably decreases as the sweep rate increases. This feature is conspicuous especially below 30 K (Fig. 4**b**), highlighting the kinetic aspect of the FOT process. Using the KAI model, the field-induced transition at various field-sweeping rates at 20 K can be calculated and compared with experimental results (Figs. 4**c** and 4**d**). As in the case of the temperature dependence of the hysteresis line, we find that the normalized sweep-rate dependence [$\delta H_c(r)$–$\delta H_c(r_{\min})$]/[$\delta H_c(r_{\max})$–$\delta H_c(r_{\min})$] depends on $N$ and $n$ only weakly. We therefore compare this value between the experiments and calculations at selected temperatures, as shown in Fig. 4**e** (see also Supplementary Figs. 4**d-f**), confirming that the sweep-rate dependence of the transition field is successfully captured by the calculation at a quantitative level.

**Conclusion**

The present analysis relies on extrapolating the $v(T, \delta H)$ data towards $H_{\text{low}}$ within the framework of the modified Merz's law (equation (1)), and thus, the validity of the model near $H_{\text{low}}$ may not be necessarily clear. However, we have demonstrated that the two main characteristics of the hysteresis line at low temperatures, that is, the pronounced broadening and the pronounced sweep-rate dependence are both quantitatively explained by considering the creep motion of the DW obeying an activate form with respect to temperature and magnetic field. These observations indicate that the modified Merz's law still captures the DW dynamics near $H_{\text{low}}$. Given that not all FOT materials exhibit pronounced hysteresis broadening, the DW creep dynamics, including the $\mu$ value, are likely to vary widely from material to material. So far, DW dynamics has been investigated exclusively in ferroic systems because of their importance in applications. Toward exotic applications based on an FOT beyond ferroic systems, our findings suggest that control of $\mu$ in a phase-competing system is likely a key ingredient.

## Methods

**Crystal growth.** Single crystals of $(Fe_{0.95}Zn_{0.05})_2Mo_3O_8$ were grown by chemical vapor transport reaction method following the literature[24,25,41]. Powders of $MoO_2$, Fe, $Fe_2O_3$, and ZnO were mixed and ground in the stoichiometric molecular ratio. The powder was sealed in the inner quartz tube (diameter 15 mm, length 130 mm) together with 100 mg of $TeCl_4$ as the transport agent. This ampoule was double-sealed with the outer quartz tube (diameter 20 mm, length 200 mm), and placed in a three-zone furnace. After 24 hours of back transport (845 °C on the precursor-loaded side; 980 °C on the growth side), the temperature gradient was reverted to be 980 °C for the load side and 845 °C for the growth side, and maintained for 11 days.

**Magnetization measurement.** Magnetization was measured by reciprocating sample option (RSO) and vibrating sample magnetometer (VSM) modes in a commercial superconducting quantum interference device magnetometer (MPMS-XL and MPMS-3; Quantum Design). Magnetic sweeping rates from $2 \times 10^{-4}$ T s$^{-1}$ to $2 \times 10^{-2}$ T s$^{-1}$ were controlled in MPMS-3 and the VSM measurement was performed in the continuous sampling mode.

**Magnetic force microscopy measurement.** Frequency-modulated MFM was performed under high vacuum conditions with a commercially available scanning probe microscope (attocube AFM/MFM I). We used the MFMR tip (supplied by NANOSENSORS) with a tip radius of ~50 nm. The typical sample dimension was approximately 3.0×2.0×1.0 mm$^3$ with a mass of 31.95 mg. We prepared a sample with a naturally grown surface that is normal to the *c*-axis. Gold with a thickness of 50–100 nm was deposited on the surface to avoid the charge up problem during the measurement. The MFM measurement was performed in noncontact mode with a lift height of 100–200 nm. The amplitude of the cantilever oscillation was approximately 10–20 nm. The resonant frequency was ≈63.6 kHz, and the Q-factor was ≈3.5×10$^4$ under the measurement conditions.

**Magneto-optical Kerr effect (MOKE) measurement.** MOKE measurement setup follows the literature[42]. A commercial polarizing microscope (BXFM, Olympus) was fixed to the aluminum frame self-constructed above the physical property measurement system (PPMS; PPMS-14T, Quantum Design). An infinity-corrected objective lens (PLN4XP, Olympus) attached to a fiberglass reinforced plastic tube (diameter 16 mm, length 1m) was inserted into the sample chamber of PPMS. Light from a LED source (625 nm, M624L4-C1, Thorlabs) passed through a polarizer and coaxially illuminated onto the sample. Reflected light through the objective lens, an analyzer, and a television (TV) lens (magnification of 5×) was captured by a CMOS camera (PCO). The polarization is shifted about 5 degrees from the orthogonal configuration[42]. The temperature and magnetic field were controlled by the PPMS control system. The sample on the sapphire plate was placed on the PPMS sample puck (P101/3A) with varnish and installed into the sample chamber of PPMS. The measurement was performed with the naturally grown surface of the same sample used in the MFM measurement.

## Data availability

The data are available from the corresponding authors upon reasonable request.

## Code availability

The source codes are available from the corresponding authors upon reasonable request.

**Author contributions**

K.M. and F.K. planned the project. K.M. and Y.N. performed the magnetization measurement and the magnetic force microscopy measurement, and analyzed the data. K.M. and Y.K. performed the MOKE experiment and K. M. analyzed the data. T.K. synthesized the single crystals used for this study. K.M. performed the KAI model calculation. K.M. and F.K. wrote the manuscript. All authors discussed the results and commented on the manuscript.

**Acknowledgments**

This work was supported by JSPS KAKENHI (Grant No. 21K14398, No. 21H04442, and No. 18H05225, No. 22H01164, No. 23H04862), JST PRESTO (No. JPMJPR21Q2) and JST CREST (No. JPMJCR1874). K.M. was supported by the Special Postdoctoral Researcher Program of RIKEN. The crystal structure was visualized by VESTA 3[43].

**Competing interests**

The authors declare that there are no competing interests.

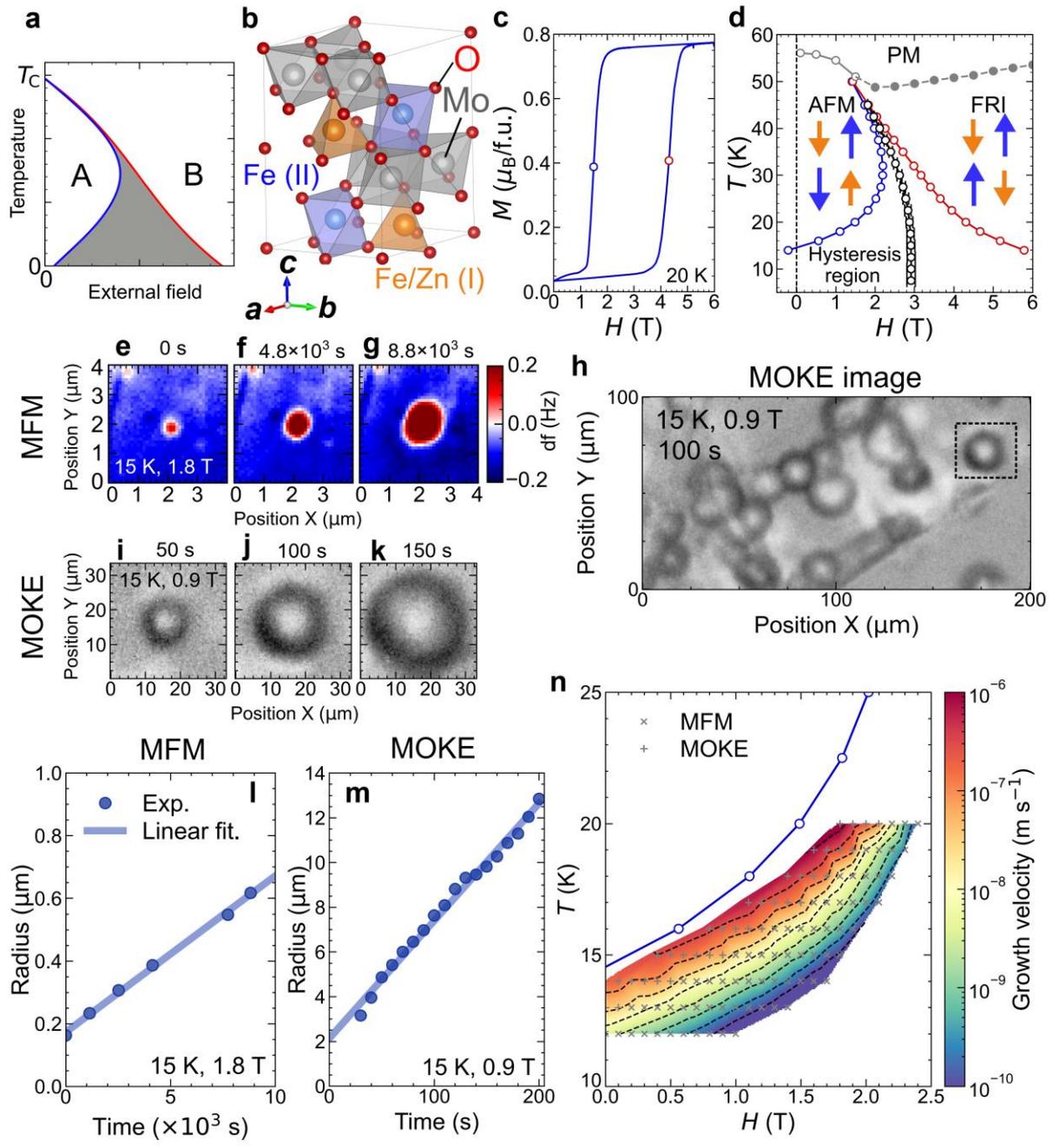

**Fig. 1 | Low-temperature hysteresis broadening and growth of the AFM domain in (Fe$_{1-y}$Zn$_y$)$_2$Mo$_3$O$_8$. a**, Archetypal phase diagram of a field-induced first-order phase transition (FOT) between two competing phases (labelled by A and B) with different symmetries. The solid lines in **a** represent magnetic fields at which the phase transition is observed during magnetic field sweeps at a given rate. Such hysteresis broadening is not always but is frequently observed in a field-induced FOT between competing phases. **b** The crystal structure of (Fe$_{1-y}$Zn$_y$)$_2$Mo$_3$O$_8$. Fe$^{2+}$ ions at the Fe(I)-sites (brown) and Fe(II)-sites (blue) are surrounded by oxygen tetrahedra and octahedra, respectively. Mo$^{4+}$ ions (gray) form the nonmagnetic spin-trimer. **c** Isothermal magnetization curve of (Fe$_{0.95}$Zn$_{0.05}$)$_2$Mo$_3$O$_8$ at 20 K. Blue (red) open circles indicate the transition fields from the FRI to AFM (AFM to FRI) phases, where AFM and FRI represent antiferromagnetic and ferrimagnetic, respectively. These transition fields are defined as the inflection points of *M-H* curves. **d** Temperature-magnetic-field phase diagram of (Fe$_{0.95}$Zn$_{0.05}$)$_2$Mo$_3$O$_8$. The transitions from the AFM to FRI phases and from the FRI to the AFM phases are indicated by red and blue open circles, respectively. Black open circles with error bars represent the equilibrium FOT phase boundary, taken from ref. [32]. Orange and blue arrows in the schematic represent spins at Fe(I) and Fe(II) sites, respectively. **e-g** Magnetic force microscopy (MFM) images of a time-evolving AFM domain at 15 K and 1.8 T at (**e**) 0 s, (**f**) 4.8×10$^3$ s and (**g**) 8.8×10$^3$ s. **h** Magneto-optical-Kerr-effect-microscopy (MOKE) image of a wider area at (15 K, 0.9 T). **i-k** MOKE images of a time-evolving AFM domain at (15 K, 0.9 T) indicated by the dotted rectangle in panel **h** at (**i**) 50 s, (**j**) 100 s and (**k**) 150 s. Images shown in panels **h-k** are divided by the raw MOKE image at 0 s to remove the influence of the position-dependent intensity of the incident light. **l,m** Time dependences of a radius at (**l**) (15 K, 1.8 T) and (**m**) (15 K, 0.9T). **n** Contour map of the growth velocity, superimposed on the phase diagram. The blue solid line with blue open circles indicates the transitions from the FRI to AFM phases on a field-decreasing process. The black dashed lines on the contour map represent contour lines drawn at 1 and 3 series from 1.0×10$^{-10}$ to 1.0×10$^{-6}$ in logarithmic representation. Gray cross and plus marks represent the points measured with MFM and MOKE, respectively.

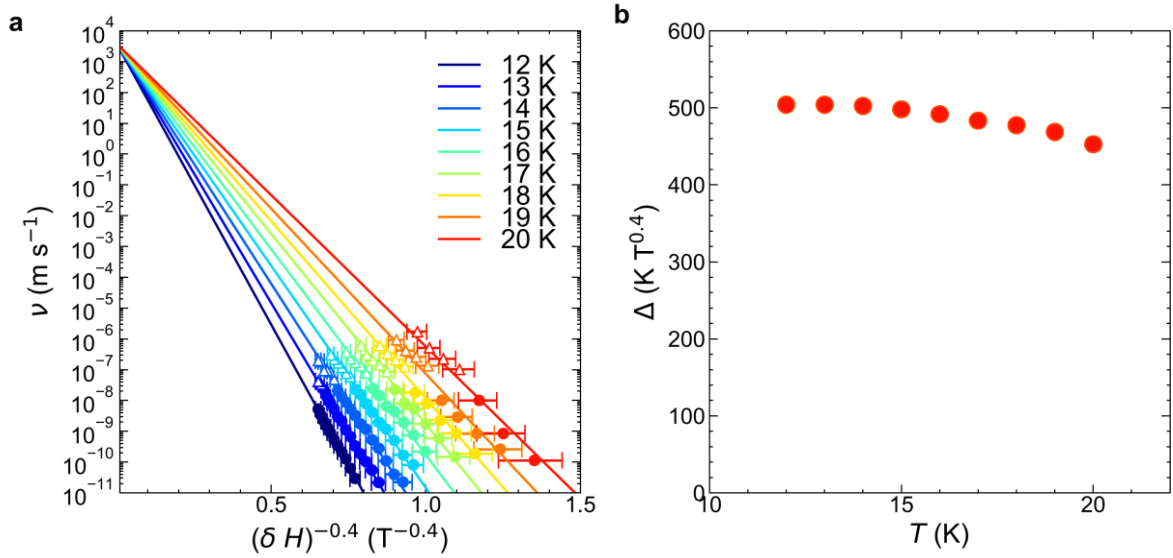

**Fig. 2 | The fitting of growth velocities of an antiferromagnetic domain with Merz's law. a** Growth velocities plotted against the inverse of the driving magnetic field. The data indicated by filled circles and open triangles are obtained from the MFM and MOKE experiments, respectively. The solid lines represent the fitting with $v(T, \delta H) = A \exp\left[-\frac{\Delta}{T \times (\delta H)^\mu}\right]$ with $\mu$ = 0.4, where δ$H$ represents the driving force of the phase transition and it is given by $H - H_0$ with $H_0$ being the equilibrium FOT phase boundary. The error bars of the data points originate from the error bars of the thermodynamic phase boundary displayed in Fig. 1**d**. **b** Temperature dependence of the activation energy Δ in equation (1), obtained from the fitting in (**a**).

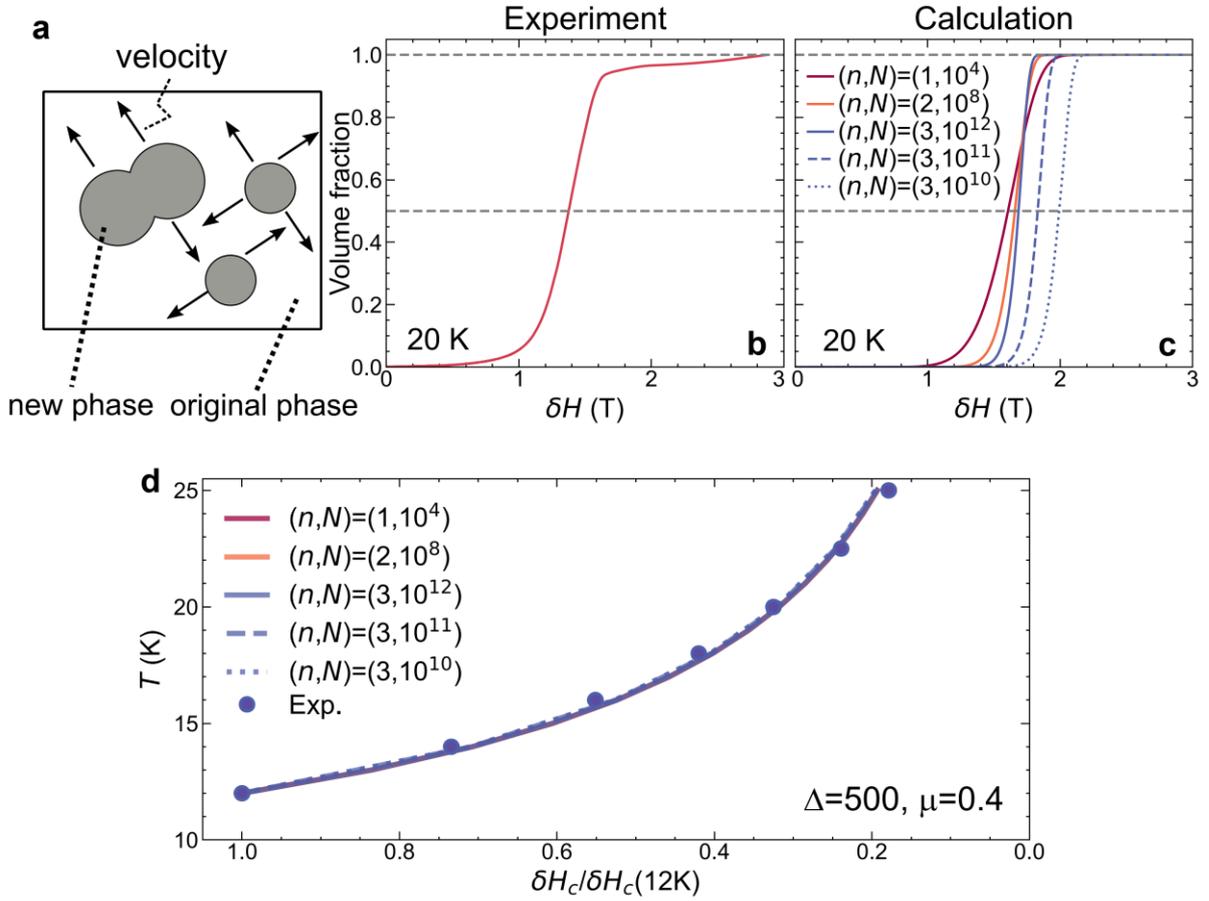

**Fig. 3 | Calculation of the low-temperature hysteresis broadening based on the Kolmogorov-Avrami-Ishibashi (KAI) model combined with modified Merz's law.**
**a** Schematic illustration of the KAI model for a two-dimensional case. Gray regions represent nuclei of a new phase, which are isotropically growing. **b,c** The experimental (b) and corresponding calculated (c) transition profiles at 20 K. The vertical axis in panel **b** represents the volume fraction that is calculated as [$M(7T)$–$M(H)$]/[$M(7T)$–$M(0T)$], where $M(H)$ is magnetization. The horizontal axis is $\delta H(T) = |H(T) - H_c(T)|$. The calculation was performed following equation (3) at selected values of ($n,N$), where $n$ and $N$ represent the system dimension and the density of the nucleation sites, respectively. **d** Comparison of the transition fields from the FRI to AFM phases, obtained from the calculation and experiments. The transition fields are normalized by the transition field at 12 K. The experimental and simulated data are represented by the filled circles and curves, respectively.

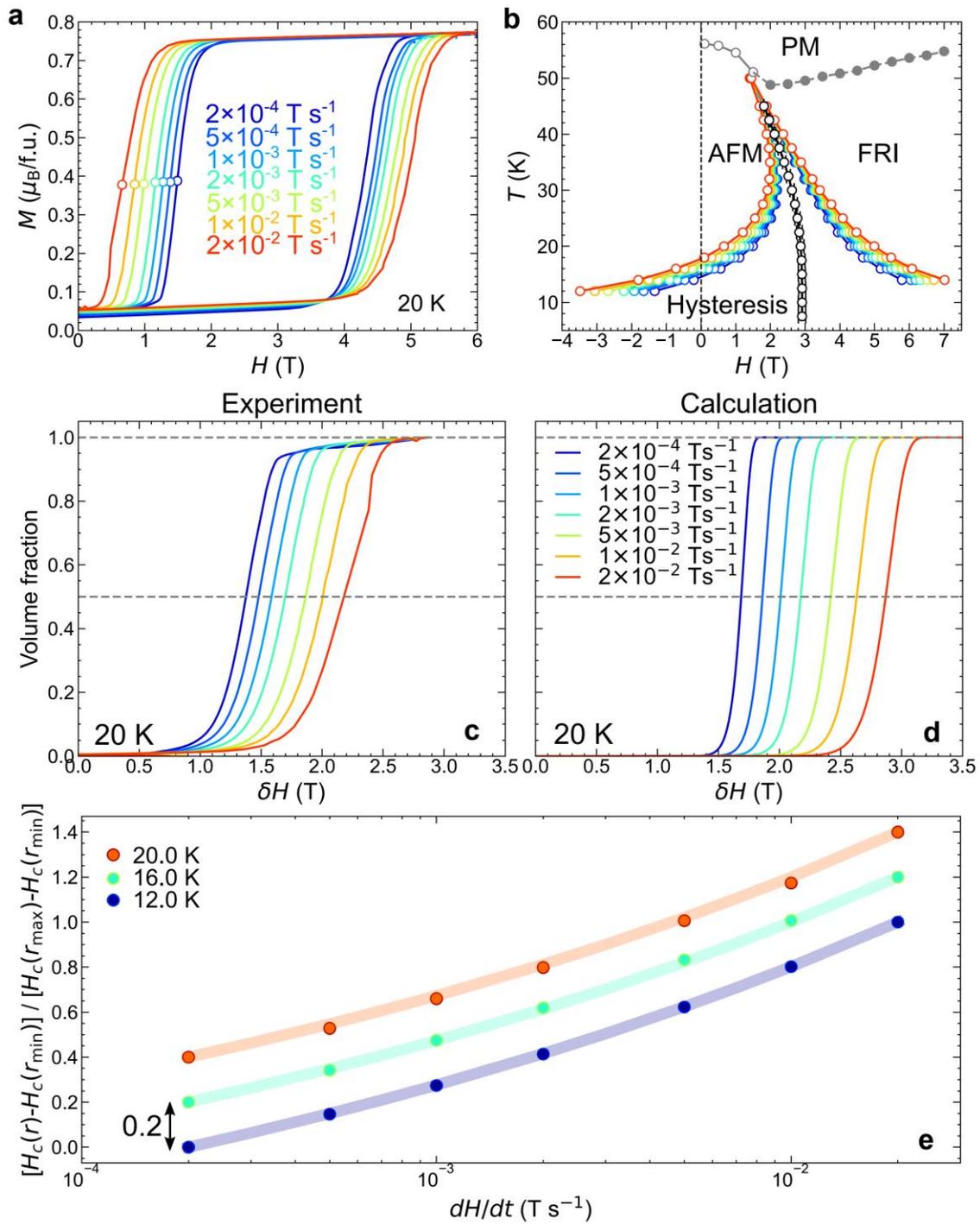

**Fig. 4 | Sweep-rate-dependent behaviour of the hysteresis lines. a** Isothermal magnetization curves at 20 K measured at various sweep rates, obtained in the experiments. **b** The transition lines determined by various sweep rates. Open circles represent the sweep-rate dependent transition fields from FRI (AFM) to AFM (FRI) phases, determined by isothermal magnetization-curve measurements. **c,d** The experimental (c) and corresponding calculated (d) sweep-rate-dependent transition profiles at 20 K. The definitions of the vertical and horizontal axes are the same as in Figs. 3**b** and 3**c**. **e** Comparison of the sweep-rate dependent transition fields from the FRI to the AFM phases at selected temperatures, obtained from the calculation and the experiments. The transition fields are normalized as $[\delta H_c(r) - \delta H_c(r_{min})]/[\delta H_c(r_{max}) - \delta H_c(r_{min})]$ at the lowest and highest sweep rates, i.e., $r_{min} = 2\times10^{-4}$ T s$^{-1}$ and $r_{max} = 2\times10^{-2}$ T s$^{-1}$. The experimental and simulated data are represented by the filled circles and curves, respectively. Each curve is shifted by 0.2 for visibility.

# Supplementary Information for

**Low-temperature hysteresis broadening emerging from domain-wall creep dynamics in a two-phase competing system**

Keisuke Matsuura[1], Yo Nishizawa[2], Yuto Kinoshita[3], Takashi Kurumaji[4], Atsushi Miyake[3], Hiroshi Oike[1,2,5], Masashi Tokunaga[1,3], Yoshinori Tokura[1,2,6] and Fumitaka Kagawa[1,2,7]

[1] *RIKEN Center for Emergent Matter Science, Wako 351-0198, Japan*
[2] *Department of Applied Physics and Quantum-Phase Electronics Center (QPEC), University of Tokyo, Tokyo 113-8656, Japan*
[3] *The Institute for Solid State Physics, The University of Tokyo, Kashiwa, Chiba 277-8581, Japan*
[4] *Department of Advanced Materials Science, University of Tokyo, Kashiwa 277-8561, Japan*
[5] PRESTO, Japan Science and Technology Agency (JST), Kawaguchi 332-0012, Japan
[6] *Tokyo College, University of Tokyo, Tokyo 113-8656, Japan*
[7] *Department of Physics, Tokyo Institute of Technology, Tokyo 152-8551, Japan*

## Supplementary Note 1: Identification of the AFM domain with MFM

Supplementary Fig. 1 compares the magnetic force microscopy (MFM) and topographic images to show whether the observed circular domain represents the AFM domain. The dark circular area seen at approximately (X, Y) = (4 μm, 5 μm) in the MFM image (Supplementary Fig. 1a) is not discernible in the topographic image (Supplementary Fig. 1b). Furthermore, considering the facts that the frequency shift, $df$, of the dark circular area is relatively positive with respect to the matrix FRI phase and that the area (Figs. 1e-g in main text) is time dependent, the observed circular area is safely assigned to an AFM domain.

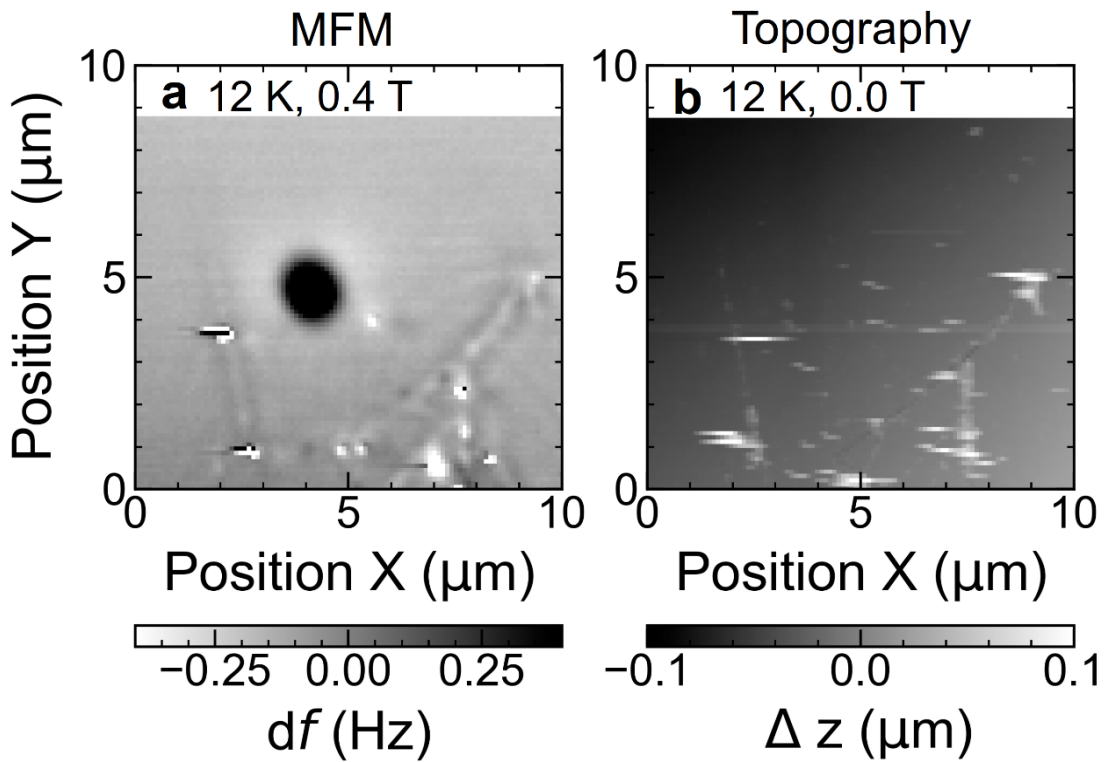

**Supplementary Figure 1 | Comparison of a magnetic force microscopy (MFM) and topographic images. a** MFM image at 12 K and 0.4 T. **b** Topographic image at 12 K and 0 T.

**Supplementary Note 2: Growth velocities at different positions**

Supplementary Fig. 2 compares the growth velocities of two different AFM domains measured with MFM. The growth velocities are in good agreement with each other. Furthermore, from the MOKE experiment, we also checked the growth velocities of five different positions. The growth velocities of each domain at 15 K and 0.9 T were 0.0599 μm s$^{-1}$, 0.0484 μm s$^{-1}$, 0.0597 μm s$^{-1}$, 0.0544 μm s$^{-1}$, and 0.0509 μm s$^{-1}$, respectively. Hence, the standard deviation estimated from the five different positions was only 0.005 μm s$^{-1}$, which is sufficiently small compared with the growth-velocity change induced by the present magnetic-field step (0.1 T).

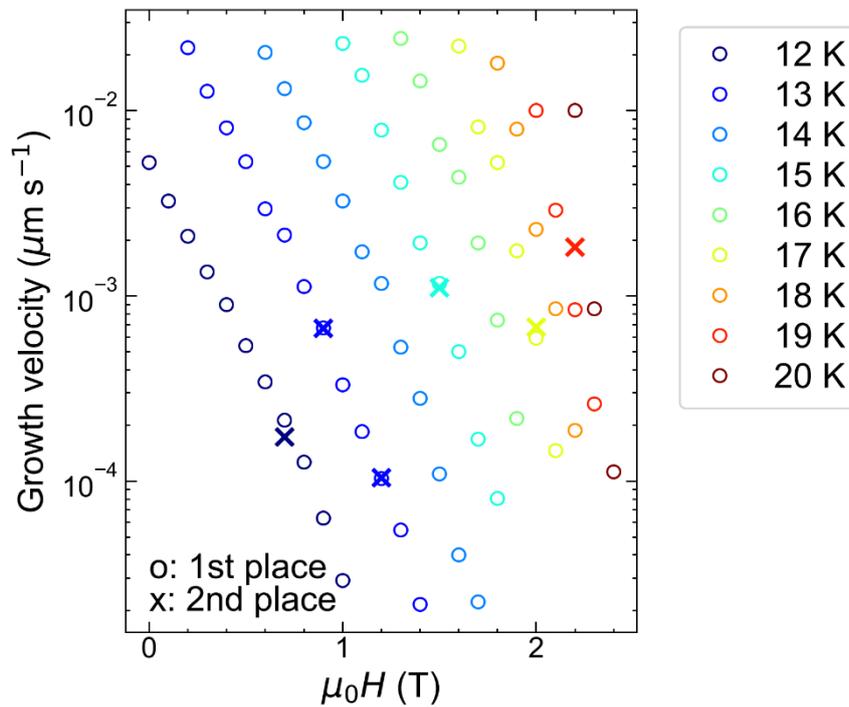

**Supplementary Figure 2 | Comparison of the growth velocities at two different positions measured by MFM.** Open circles and crosses represent the growth velocities at the 1st and 2nd places, respectively. The data displayed in the main text are those measured at the 1st place.

**Supplementary Note 3: Definition of a driving force in Merz's law**

For a ferromagnetic domain wall, it was sufficient to define the driving force with respect to a zero external field; however, this is not the case for the present system. The position of the thermodynamical equilibrium FOT line in the phase diagram should be accurately determined because the driving force originates from the difference in free energies between the AFM and FRI phases. Recently, the equilibrium FOT line has been determined experimentally in this system[S1] (white circles in Fig. 1**d**). Thus, the driving force $\delta H(T)$ at given temperature $T$ is defined as $\delta H(T) = H(T) – H_c(T)$, where $H_c(T)$ represents the equilibrium-FOT transition field at $T$.

At the thermal equilibrium transition temperature and magnetic field, the driving force is zero by definition. The difference in Gibbs free energy between the two phases is given as

$$\Delta G(T_0, H) = \Delta M^*(T_0) \times (H - H_c) + o((H - H_c)^2),$$

where $\Delta M^*(T_0)$ is the difference in magnetization between the AFM and FRI phases at $T=T_0$ and $H=H_c$. Magnetizations in the AFM and FRI phases are weakly dependent on the magnetic field[S1], and hence the second term can be neglected. In this way, as long as the regime of the weak driving force is considered (this is also essential for the justification of the creep dynamics), the driving force of the phase transition is approximated by $\propto H–H_c$.

**Supplementary Note 4: Types of disorder: random-bond and random-field**
**(4-1)  Summary of previous studies: roles of an impurity in a ferroic system**

In ferroic materials, it is known that the value of $\mu$ depends on both the dimensionality of the DW and the type of disorder, which is generally classified into random-bond (RB) or random-field (RF) types. Let us consider an Ising magnet under the influence of the impurities, with the following Hamiltonian[S2]:

$$H = -J \sum_{\langle i,j \rangle} S_i \cdot S_j + H_{RB} + H_{RF}, \quad (S1)$$

where $H_{RB}$ and $H_{RF}$ represent the Hamiltonian describing the interaction with RB- and RF-type impurities, respectively. An RB-type impurity causes the modulation of a local exchange interaction; thus, the $H_{RB}$ is given by the following Hamiltonian;

$$H_{RB} = -\sum_{\langle i,j \rangle} \Delta J_{ij}\, S_i \cdot S_j, \quad (S2)$$

where $\Delta J_{ij}$ represents the modulation of the magnetic interaction between the $i$ and $j$ sites. An RB-type impurity does not stabilize one of the two states; that is, the local symmetry of the order parameter is maintained even in the presence of an RB-type impurity. In contrast, an RF-type impurity lifts the degeneracy between the positive and negative order parameters, <S>, and thus, $H_{RF}$ is given by the following Hamiltonian;

$$H_{RF} = \sum_{i} \Delta h_i\, S_i. \quad (S3)$$

where $\Delta h_i$ represents the modulation of the local field acting on the magnetic moment at the site $i$.

Although the above expressions are represented in the form of the magnetic Hamiltonian[S2], the interactions induced by RB- and RF-type impurities can also be expressed in terms of the elastic energy at the interface[S3,S4]. Within a linear response regime, the relative energy of a distorted interface is given by

$$H_0 = \int d^{d-1}\mathbf{x}\, \left[\frac{1}{2}\Gamma(\nabla z(\mathbf{x}))^2 + V(\mathbf{x}, z(\mathbf{x}))\right], \quad (S4)$$

where the first and second terms represent an elastic energy induced by the distortion from a reference flat plane with stiffness $\Gamma$ and an interaction energy at the interface with randomly distributed impurities at $(\mathbf{x}, z(\mathbf{x}))$, respectively. $\mathbf{x}$ and $z(\mathbf{x})$ represent the coordinate at the interface and a displacement in the direction perpendicular to the flat reference interface (Supplementary Fig. 3**a**), respectively. Without $V(\mathbf{x}, z(\mathbf{x}))$, an interface favour a flat surface (black dashed line in Supplementary Fig. 3a). In the presence of $V(\mathbf{x}, z(\mathbf{x}))$, by contrast, an interface can be distorted from a flat surface to gain potential energy (red solid line in Supplementary Fig. 3a). In the strong pinning case, the interaction potentials of RB- and RF-type impurities are represented as follows [S3]:

$$V_{RB} = \sum_j v_j a \delta_{\mathbf{x}}(\mathbf{x} - \mathbf{x_j}) \delta_z(z(\mathbf{x}) - z_j), \tag{S5}$$

$$V_{RF} = \sum_j v_j \delta_{\mathbf{x}}(\mathbf{x} - \mathbf{x_j}) \int_0^{z(\mathbf{x})} dz' \delta_z(z' - z_j), \tag{S6}$$

where $a$ represents the range of interaction with respect to the direction perpendicular to the interface and $v_j$ represents the potential energy when an interface interacts with the impurity at site $j$. As expressed by Eq. (S5), an RB-type impurity contributes to the energy only when the interface just hits the impurity position; in this sense, an RB-type impurity is often referred to as short-ranged[S4]. In contrast, as expressed by Eq. (S6), the interaction potential for an RF-type impurity is given by the integral from the bulk (inside) in one phase to the interface. This means that an RF-type impurity contributes to the energy as long as it is included in bulk phases; as such, an RF-type impurity is often referred to as long range [S4]. The impacts of the two types of impurities on the energy landscape in a ferroic system[S5] are schematically shown in Supplementary Figs. 3b and 3c.

Next, we explain a dynamical exponent $\mu$, which depends on the type of impurities. $\mu$ is represented as follows:

$$\mu = \frac{d - 2 + 2\xi}{2 - \xi}, \tag{S7}$$

where $d$ is the dimensionality of a deformable object ($d$ = 1, 2 and 3 for single flux lines, surfaces or interfaces, and charge density waves or flux line lattices) and $\xi$ is a so-called roughness exponent representing the extent of the roughness of a domain wall. $\xi$ can be experimentally extracted from the square of the relative displacements between $z(\mathbf{x})$ and $z(\mathbf{x}+L)$[S3], i.e., $<|z(\mathbf{x}) - z(\mathbf{x}+L)|^2>$, which scales with the length L as $\sim L^{2\xi}$. Theoretically, the roughness exponents $\xi_{RB}$ and $\xi_{RF}$ for domain walls under the influence of either RB- or RF-type impurities are given by

$$\xi_{RB} = \begin{cases} \frac{2}{3} \text{ (for } d = 1) \\ 0.2084(4 - d) \text{ (for } d = 2, 3) \end{cases}, \tag{S8}$$

and

$$\xi_{RF} = \frac{4-d}{3} \text{ (for } d = 1, 2, 3), \tag{S9}$$

respectively[S6]. By substituiting these $\xi$ values into equation (7), the $\mu$ values for both types of impurities can be obtained. For the random-bond scenario, $\mu_{RB}$ is 0.25, 0.53, and 0.79 for $d=1$, 2, and 3, respectively. For the random-field scenario, $\mu_{RF}$ is 1, irrespective of $d$. Therefore, one can classify the types of an impurity into either an RB or an RF type, by referring to whether a dynamical exponent $\mu$ deviates from 1; this criterion can often be seen in previous studies on domain-wall dynamics in a ferroic system.

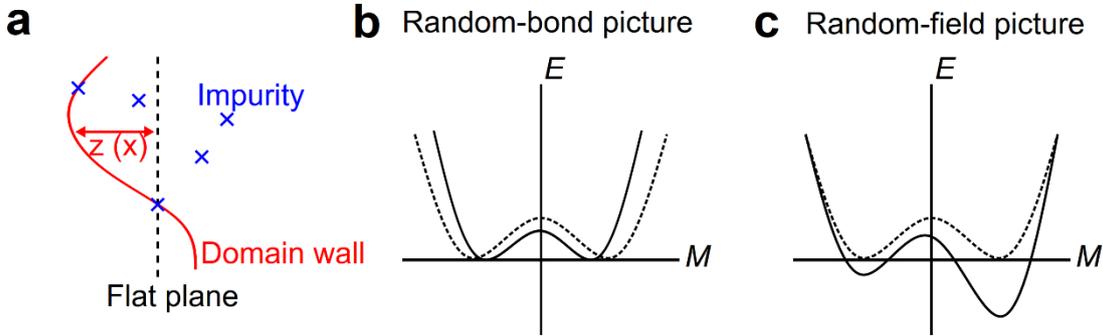

**Supplementary Figure 3 | Schematic pictures of the effects induced by random-bond and random-field impurities. a** Schematic illustration of the coupling between a domain wall (red solid line) and impurities (blue cross mark). The dashed line represents a reference flat domain wall plane without any impurity. **b,c** Impact of random-bond type (**b**) and random-field type (**c**) impurities on the energy landscape in a ferroic system. In both panels, the dashed lines represent the defect-free cases, and the solid lines represent the defect-present cases. These schematic figures are based on Ref. S5.

**(4-2) Effects of doping in a phase competition case: $(Fe_{1-y}Zn_y)_2Mo_3O_8$**

In the case of $(Fe_{1-y}Zn_y)_2Mo_3O_8$, judging from the fact that the obtained $\mu$ is ≈0.4, one may think the impurity should be classified as RB-type. However, if one assumes that the impurity is a doped Zn ion, then this Zn ion is expected to work as both RB-type and RF-type impurities. Here, two aspects of the doped Zn ion should be considered. First, selective occupation of Zn ions on Fe(I) sites weakens Fe(I)-Fe(I) magnetic interactions[S7] because this effect is considered to modulate the exchange interaction. Furthermore, the nonmagnetic ion is expected to reduce the energy cost of the magnetic domain wall when the magnetic domain wall is on nonmagnetic impurity sites. Thus, doped Zn ions should work as RB-type impurities in the present system. Second, given that a delicate energy balance between the AFM and FRI phases results from the magnitudes of the interlayer Fe(I)-Fe(II) and interlayer Fe(I)-Fe(I) interactions in this system, doped nonmagnetic Zn ions are expected to stabilize the FRI phase by weakening Fe(I)-Fe(I) magnetic interactions. In fact, as the amount of Zn doping increases, the FRI phase tends to be realized at relatively low magnetic fields, indicating that Zn doping plays a role in changing the energy balance between the two competing phases. These two mechanisms are expected to work simultaneously, and therefore, a doped Zn ion can be considered to have characteristics of both RF-type and RB-type impurities.

For a system that exhibits competition between phases with different symmetries, any impurity can have an asymmetric impact on the two competing phases, and it can also interact with the domain wall when the domain wall is on the impurity. It thus appears that any impurity can invariably have the characteristics of RB- and RF-type impurities, and it is therefore nontrivial what value of µ should be obtained by the introduction of impurities in a phase-competing system, such as $(Fe_{1-y}Zn_y)_2Mo_3O_8$.

**Supplementary Note 5: Kolmogorov-Avrami-Ishibashi (KAI) model simulation**

The Kolmogorov-Avrami-Ishibashi (KAI) model can be used to relate a microscopic domain growth velocity to the volume fraction of a growing domain while avoiding the domain coalescence problem. Solutions are given for two main physical mechanisms: continuous nucleation and one-step nucleation[S8]. The continuous nucleation mechanism (category I) represents the process in which the nucleation rate is spatially homogeneous and continuously occurs with a constant rate during the whole transformation. In contrast, in the one-step nucleation mechanism (category II), all nuclei of the new phase are created at the beginning of the phase evolution, and the subsequent domain growth evolves without being accompanied by any further nucleation. In the present study, we applied the formalism of the KAI model for the one-step nucleation mechanism because we found that nucleation always occurs at the same positions and the same magnetic field for repeated experiments, consistent with the behaviour considered in Category II.

**(5-1) Hysteresis broadening**

First, we investigate the impacts of the growth dimensionality $n$ and the nucleus density $N$ on the hysteresis broadening, because these two parameters are inherent to the KAI model. In the following calculations, we fix $\Delta = 500$ in equation (1). While the growth dimensionality is fixed to $n=3$, we perform the calculations for various values of $N$; we thus find that the hysteresis broadening does not depend on $N$, as shown in Supplementary Fig. 4**a**. Next, we perform calculations with different growth dimensionalities and confirm that the dependencies of the broadening on $n$ are very weak (Supplementary Fig. 4**b**). Therefore, we conclude that the hysteresis broadening arises from Merz's law-type, thermally activated creep motion. Finally, we examine how the dynamical exponent $\mu$ affects the hysteresis broadening. As the value of $\mu$ increases, the hysteresis broadening becomes weaker (Supplementary Fig. 4**c**). These simulation results indicate that $\mu$ strongly affects the hysteresis broadening for a given $\Delta$, although the values of $\Delta$ might also affect the hysteresis broadening.

**(5-2) Sweep-rate dependence**

We have also investigated the sweep-rate dependence of the transition fields. As shown in Supplementary Figs. 4**d**-**f**, the experimental results are well reproduced by using three different parameter sets, indicating that the normalized sweep-rate dependence is almost irrelevant to $n$ and $N$.

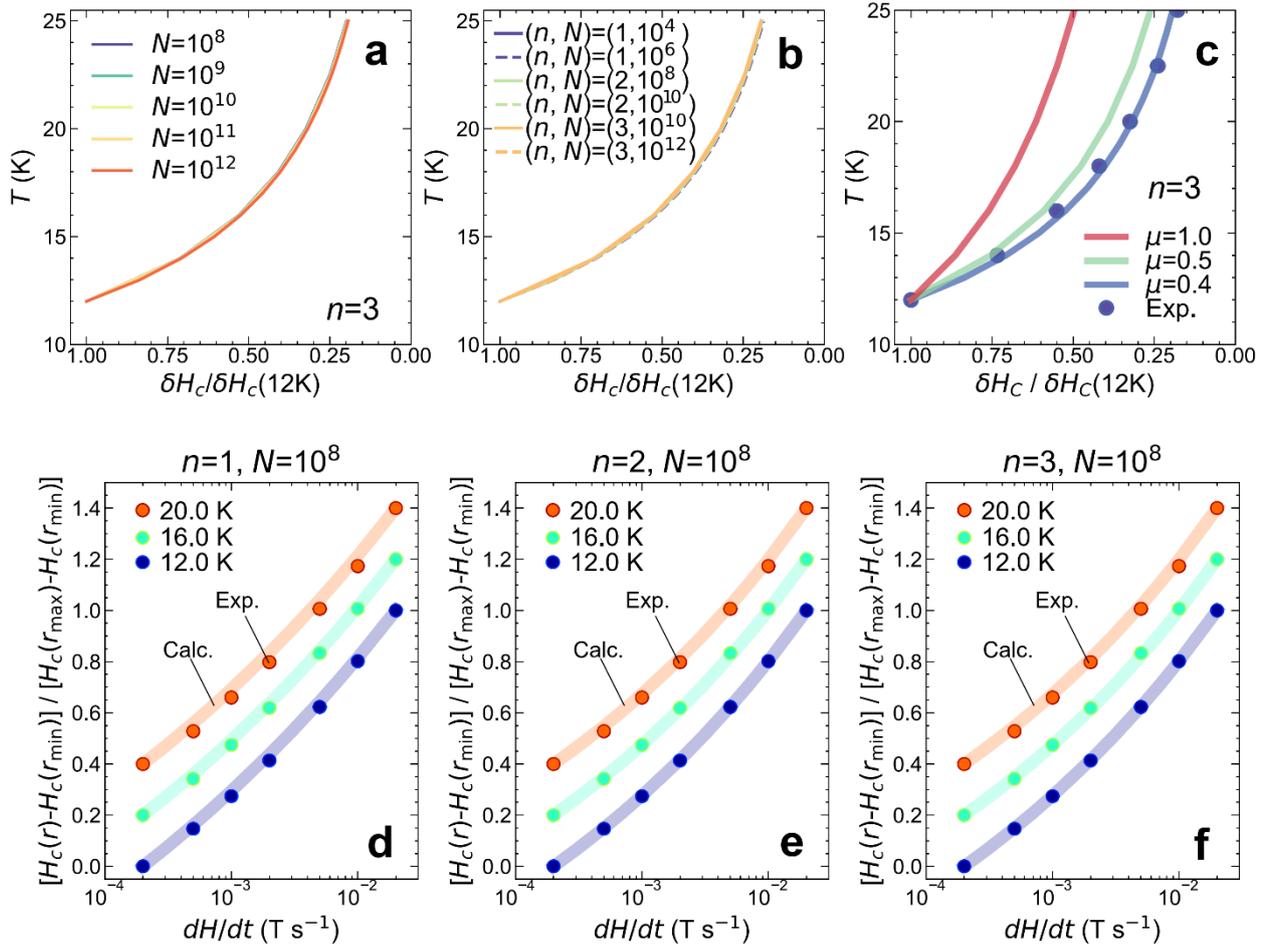

**Supplementary Figure 4 | KAI model simulation. a-c** Variations in the temperature dependences of the normalized transition fields when changing the nucleation density $N$ (**a**), growth dimensionality $n$ (**b**), and dynamical exponent $\mu$ (**c**). **d-f** Sweep-rate dependent transition fields from the FRI to the AFM phases at selected temperatures (12 K, 16 K, and 20 K) for three different parameter sets; $(n, N) = (1, 10^8)$, $(2, 10^8)$, and $(3, 10^8)$ for (**d**), (**e**), and (**f**), respectively. (**f**) is the same panel as Fig. 4e in the main text. The transition fields are normalized as $[\delta H_c(r) - \delta H_c(r_{min})]/[\delta H_c(r_{max}) - \delta H_c(r_{min})]$ at the lowest sweep rate, $r_{min} = 2 \times 10^{-4}$ T s$^{-1}$ and $r_{max} = 2 \times 10^{-2}$ T s$^{-1}$. Corresponding calculated curves at 12 K, 16 K, and 20 K are indicated by blue, green, and orange solid lines, respectively. Each curve is vertically shifted by 0.2 for visibility.